\documentclass{elsart}
\usepackage{natbib}
\begin{document}
\runauthor{}
\begin{frontmatter}
\title{Constraints on Inflation}
\author{Pedro T. P. Viana}
\address{Centro de Astrof\'{\i}sica da Universidade do Porto,
Rua das Estrelas s/n, 4150 Porto, Portugal, and Departamento de Matem\'{a}tica 
Aplicada da Faculdade de Ci\^{e}ncias da Universidade do Porto; viana@astro.up.pt}
\begin{abstract}
A short introduction to structure formation is given, followed by a discussion of 
the possible characteristics of the initial perturbations assuming a 
generic inflationary origin. Observational data related 
to large-scale structure and the cosmic microwave background radiation is then used 
in an attempt to constrain the characteristics of such perturbations. Future directions  
are also explored. 

The possibility of direct detection of a stochastic gravitational wave background produced 
during an inflationary phase in the early Universe is briefly discussed, as well as 
the available evidence regarding the present value of the total energy density in the Universe.

\end{abstract}
\end{frontmatter}

\section{Overview}

The simplest models of inflation, those in which only one scalar 
field is present and is minimally coupled, lead to very simple and 
clear predictions: (1) the Universe is simply connected, (nearly) homogeneous 
and isotropic, with no detectable large-scale rotation, at least up to 
scales slightly larger than the present particle horizon; (2) the 
observable Universe is spatially flat; (3) the scalar perturbations that 
eventually originated the large-scale structures observed today, if generated 
during inflation, were primordial (i.e. passive), adiabatic and Gaussian 
distributed, with a nearly scale-invariant power spectrum.

However, in more complicated models, where more than one scalar field is present, 
with the possibility of the such fields being strongly coupled, most of the above 
predictions can be weakened. The exceptions are those of a trivial topology, the 
absence of large-scale rotation, and the primordial nature of the scalar perturbations. 
Nevertheless, these are still sufficient to provide tests for both hypothesis: of 
inflation as the event responsible for the present-day large-scale Universe being 
nearly homogeneous, isotropic and spatially flat; and of inflation as the most 
important mechanism behind the generation of the scalar perturbations that eventually 
originated the large-scale structures observed in the Universe today.

I will focus on the later hypothesis [see \cite{LiddleCarg} for a general review of inflation], 
describing the observational 
tests, and present constraints, on the various characteristics expected for scalar perturbations in 
the simplest inflationary models. These characteristics have an impact both on the nature and 
timescale for the formation and evolution of structures on large-scale in the Universe, and on the 
properties of the cosmic microwave background radiation (CMBR). I will therefore 
spend most of this review on how the present observational data regarding these two 
topics constrain the characteristics of the scalar perturbations, and what in turn that tells us 
about inflation. I will also discuss the prospects of detecting locally a possible stochastic 
background of gravitational waves produced during inflation, and the present evidence regarding the 
geometry of the Universe, and whether it supports the prediction of spatial flatness associated 
with the simplest inflationary models. Though tensor perturbations and spatial flatness cannot be 
used to test inflation, given that inflationary models exist which do not predict them, they can 
offer strong support to the hypothesis that an inflationary period did occur in the early Universe. 

\section{Structure formation and inflation}

\subsection{Introduction}

In studies of structure formation one is particularly interested in the statistical 
properties of the density contrast of the matter distribution, 

$$
\delta({\bf x},t)\equiv\frac{\rho({\bf x},t)-\overline{\rho}}{\overline{\rho}}\,, 
$$

which is defined in terms of the density field $\rho({\bf x},t)$ and the 
{\it comoving} mean background density $\overline{\rho}$, where ${\bf x}$ is the 
{\it comoving} position. 

A Fourier expansion can be made in a large enough box (of volume V) so that 
$\delta({\bf x},t)$ is periodic within the box. We then have 

$$
\delta({\bf x},t)=\sum_{k}\delta_{{\bf k}}(t)\,e^{i{\bf k}\cdot{\bf x}}\,,
$$

where $k\equiv|{\bf k}|$ is a comoving wavenumber, with the Fourier coefficients being given by 

$$
\delta_{{\bf k}}(t)=\frac{1}{V}\int\delta({\bf x},t)\,
e^{-i{\bf k}\cdot{\bf x}}d{\bf x}\,.
$$

We will define the {\it power spectrum}, ${\it P}(k,t)$, as 

$$
{\it P}(k)\equiv\left(\frac{Vk^{3}}{2\pi^{2}}\right)|\delta_{k}|^{2}\,.
$$

The dispersion of the density contrast is then simply given by 

$$
\sigma^{2}(t)\equiv\left\langle\delta^{2}({\bf x},t)\right\rangle
=\int_{0}^{\infty}{\it P}(k,t)\,\frac{dk}{k}\,.
$$

If it is assumed that for any given realization of the volume V the phases
of the Fourier coefficients $\delta_{\bf k}$ are uncorrelated, 
the central limit theorem then guarantees that at any point the 
density contrast $\delta({\bf x},t)$ obeys Gaussian statistics. The 
probability distribution of $\delta({\bf x},t)$ at each point is then 

$$
p(\delta)\,d\delta=\frac{1}{\sqrt{2\pi}\sigma}\,\exp
\left(-\frac{\delta^{2}}{2\,\sigma^{2}}\right)\,d\delta\,.
$$

This equation implies that there is always some probability of having 
$\delta({\bf x},t)<-1$, which, by definition, is not physically possible. 
Therefore, as a first approximation, it is only valid to consider 
$\delta({\bf x},t)$ as a Gaussian random field if there is
only a very small probability of having $\delta({\bf x},t)<-1$ by the 
above equation, i.e. if $\sigma(t)\ll1$. This condition is also  
necessary if we wish to use linear perturbation theory to follow the evolution 
of $\delta({\bf x},t)$. Gaussian random fields are very special, since only the 
power spectrum is required to specify all of the statistical properties of the field, 
whereas for non-Gaussian fields the full hierarchy of probability distributions is needed. 

After matter domination, the power spectrum of the density contrast, $\delta({\bf x},t)$, 
can be written as \cite{LLPR,LL,Pea}

$$
{\it P}(k,t)=\frac{g^{2}(\Omega,\lambda)}{g^{2}(\Omega_0,\lambda_{0})} \, 
\left( \frac{k}{aH}\right)^4 T^{2}(k,t)\,\delta_{H}^{2}(k) \,,
$$

where the quantities $aH$, $\Omega$ and $\lambda\equiv\Lambda c^{2}/3H^{2}$ are 
to be calculated at $t$. The function $g(\Omega,\lambda)$ accounts for the rate of 
growth of density perturbations relative to the Einstein-de Sitter case, whose growth is 
given by the $(aH)^4$ factor. The transfer function, $T(k,t)$, measures the change at $t$ 
in the amplitude of a perturbation with comoving wavenumber $k$ relative to a perturbation 
with infinite wavelength, thus in the limit $k\rightarrow0$ (in practice $k\rightarrow k_{hor}$ 
due to gauge ambiguities), one has $T(k,t)\rightarrow1$. 
The shape of the transfer function results mostly from the different
behaviour of perturbations in the radiation and matter dominated
eras, and from sub-horizon damping effects, like Silk damping,
which affects baryons, and free-streaming (Landau damping),
which acts on hot dark matter perturbations. An oscillatory pattern can also 
appear in the transfer function if baryons contribute significantly to the
matter density in the Universe, due to acoustic oscillations 
of the photon-baryon fluid on scales below the horizon until decoupling
occurs. The calculation of a transfer function not only depends on the type of mechanism 
responsible for the generation of the density perturbations, but also on the assumed matter 
and energy content in the Universe. It thus needs to be determined numerically, though 
nowadays there are several analytical prescriptions which approximate it for the most
popular structure formation scenarios [see e.g. \cite{EH}].

The quantity $\delta_{H}^2(k)$, defined as 

$$
\delta^{2}_{H}(k)\equiv
\left\langle\left(\frac{\delta\rho}{\rho}\right)^{2}\right\rangle_{aH=k}\,,
$$

specifies the power spectrum of density perturbations at horizon re-entry.
In the simplest inflationary models it can be well described by a single power-law, 

$$
\delta_{H}^2(k)=\delta_{H}^2(k_{0})\left(\frac{k}{k_{0}}\right)^{n-1}\,,
$$

where $n$ is the so-called spectral index and $\delta_{H}(k_{0})$ is a normalisation 
factor at an arbitrary comoving wavelength $k_{0}$. Since the {\it COBE} measurement of the 
amplitude of the large-angle anisotropies in the temperature of the CMBR became available, 
the value of $\delta_{H}(k_{0})$ is usually set so as to reproduce it (though some previous 
assumption has to be made regarding the contribution of
tensor perturbations, i.e. gravitational waves, to the anisotropies).  When
this is done, in the simplest inflationary models the value of
$\delta_{H}(k_{0})$  then depends essentially only on the values of $n$,
$\Omega_{0}$ and $\lambda$ [see e.g. \cite{BW}]. 
A scale-invariant, or Harrison-Zel'dovich, power spectrum corresponds to $n=1$. In general most 
inflationary models give $n\leq1$, though in some it is possible to have $n>1$.

\subsection{Adiabatic vs. entropy perturbations}
 
Perturbations in a multi-component system can be of the {\it entropy} or of the {\it adiabatic} 
types. The first correspond to fluctuations in the form of the local equation of state of 
the system, e.g. fluctuations in the relative number densities of the different particle
types present in the system, while the second correspond to fluctuations in its energy density. 
In the case of a perfect fluid composed of matter and radiation, pure 
entropy perturbations are characterised by $\delta\rho_{r}=-\delta\rho_{m}$,  
while for pure adiabatic perturbations, 
$\delta\rho_{r}/\rho_{r}=(4/3)(\delta\rho_{m}/\rho_{m})$. The entropy perturbations are also 
called {\it isocurvature}, given that the total density of the system remains homogeneous. In 
contrast, the adiabatic perturbations are also known  as {\it curvature} perturbations, as they 
induce inhomogeneities in the spatial curvature. The two types of perturbations
are orthogonal, in the sense that all other types of perturbations on 
a system can be written as a combination of both adiabatic and entropy
modes. 

On scales smaller than the Hubble radius any entropy perturbation rapidly becomes an 
adiabatic perturbation of the same amplitude, as local pressure differences, due to the 
local fluctuations in the equation of state, re-distribute the energy density. However, 
this change is slightly less efficient during the radiation dominated era than during the 
matter dominated era (and can only occur after the decoupling between photons and baryons,  
in the case of baryonic isocurvature perturbations). Causality precludes this re-distribution 
on scales bigger than the Hubble radius, and thus any entropy perturbation on these scales 
remains with constant amplitude. The end result is that initialy scale-invariant power spectra  
of adiabatic and isocurvature perturbations give rise, after matter-radiation equality, 
to power spectra of density perturbations with almost the same shape.  
 
Entropy perturbations are not affected by either Silk or 
Landau damping, contrary to adiabatic perturbations, thus potentially providing a 
means of baryonic density perturbations existing below the characteristic Silk damping 
scale after recombination (note that this could also have been achieved if there were   
cold dark matter adiabatic perturbations at such scales, with the dark matter 
necessarily being the dominant matter component in the Universe). 

However, presently the amplitude of any primordial entropy 
perturbations is severely constrained by the level of anisotropy in the temperature 
of the CMBR as measured by {\it COBE}. In the case of an Universe with critical-density and 
scale-invariant perturbations, the total anisotropy on large angular scales is six times bigger in 
the case of pure entropy perturbations than in the case of pure adiabatic perturbations, for the 
same final matter density perturbation at those scales \cite{LLPR}. We will later see that in a 
critical-density universe with initial scale-invariant adiabatic perturbations, the amplitude of the density 
perturbation power spectrum needed to generate observed structures, like rich galaxy clusters, and that 
needed to generate the temperature anisotropies measured by {\it COBE}, are roughly compatible. Therefore  
if one requires small-scale density perturbations with high enough 
amplitude to reproduce known structures in an universe with critical-density and initial 
scale-invariant isocurvature perturbations, one ends up with CMBR anisotropies on COBE scales with much 
larger amplitude than those which are measured. 

Possible ways of escaping this handicap associated with entropy perturbations are: breaking the 
assumption of scale-invariance by assuming a steeper dependence with $k$, i.e. decreasing the amount 
of large-scale power relatively to small-scale power; and decreasing the matter density in 
the Universe, i.e. assuming $\Omega_{0}<1$. However, the first possibility leads to values for 
the spectral index which are in conflict with the constraints imposed by the 
{\it COBE} data [see e.g. \cite{Hu}], while the second solution demands 
unrealistically small values for $\Omega_{0}$, well below 0.1 \cite{Burns}. 
However, combinations of these two changes, together with the introduction of more exotic forms of dark 
matter, like decaying particles \cite{Hu}, may provide working 
models purely with isocurvature perturbations. Further, isocurvature perturbations need
not have an inflationary origin, as cosmological  defect models provide
an alternative means of generating structure from isocurvature initial conditions. 

In any case, clearly though the possibility of isocurvature perturbations is not 
yet ruled out from the point of view of structure formation, it is in much worse 
shape than the hypothesis of adiabatic perturbations, only surviving by appealing to a complex 
mixture of effects in the case of an inflationary origin, or by being associated with 
topological defects.
 
\subsection{Passive vs. active perturbations}

Perturbations can be {\it passive} (i.e. primordial) or {\it active}. Passive perturbations are those 
which are generated in the very early Universe (e.g. during inflation), and henceforth 
evolve passively, being changed only by the action of cosmic expansion and gravity. This means 
that the phases of such perturbations, in a Fourier expansion of the density field, will 
remain constant once the perturbations are generated, and as long as they evolve linearly. Therefore, 
if one adds them up over time, they add up coherently. It is in this sense that passive perturbations 
are usually also called coherent. Further, as soon as these perturbations enter the particle horizon, 
the photon-baryon fluid will try to flow into the potential wells associated with the perturbations, 
setting in motion acoustic oscillations in the fluid which will be in {\it in phase}. 
Primordial perturbations, adiabatic or isocurvature, always lead to such in-phase 
(or coherent) acoustic oscillations of the photon-baryon fluid. 

On the contrary, (random) active perturbations, because 
they are constantly being produced {\it randomly} across space, tend to produce an ensemble of 
perturbations whose phases will add up over time incoherently. Such perturbations necessarily 
lead to oscillations of the photon-baryon fluid which will be out-of-phase with each other, with 
the result that their effects will cancel themselves out. Active perturbations, like 
those produced by standard topological defect models, lead in general to incoherent acoustic 
oscillations in the photon-baryon fluid (though in a few models some coherence can temporarily 
exist on scales that have just entered the Hubble radius), as they are generated almost completely at
random. However, active perturbations are not necessarily produced in such a way. Models of active 
perturbations have been proposed, though in a somewhat contrived manner, which seem to be able to produce 
perfectly coherent oscillations in the photon-baryon fluid, analogous to those produced by passive 
perturbations [see e.g. \cite{Tur}]. However, clearly, the 
discovery that the photon-baryon fluid that existed before decoupling had coherent
acoustic oscillations, would strongly boost the status of inflation as
the best available explanation for the formation of structure.

In terms of large-scale structure, these oscillations only leave a detectable signature in the power
spectrum of density perturbations if baryons comprise in excess of about 15 per cent
of the total matter content in the Universe \cite{MWP}. The main problem is that non-linear
evolution of the power spectrum washes out the signature of the baryon
induced oscillations for large 
$k$, where the power spectrum can be better determined
observationally, leaving out  basically only the two peaks on the
largest scales as a smoking gun, where observational errors are
larger due to cosmic variance. 

\subsection{Gaussian vs. non-Gaussian perturbations}

The two characteristics of the density perturbations generated in the
simplest inflationary models that could, in principle, be more easily searched for in
large-scale structure data are their initial Gaussian probability
distribution and the near scale-invariance of their power spectrum. 

In the simplest models of inflation, due to the nature of quantum fluctuations, the phases of the 
Fourier modes associated with the perturbations in the value of the scalar field are 
independent, drawn at random from a uniform distribution in the interval 
$[0,2\pi]$. Therefore, the density perturbations resulting from these perturbations
will also be a superposition of Fourier modes with independent random 
phases. From the central limit theorem of statistics it then follows that the 
density probability distribution at any point in space is {\it Gaussian}. 
As previously mentioned, this result is extremely important, since only one function, the 
power spectrum at horizon re-entry, is then required to specify 
all of the statistical properties of the initial density distribution.  

Present large-scale structure data cannot be used to say whether the 
the initial density perturbations followed a Gaussian distribution or not. 
One problem with the detection of Gaussian initial conditions is that as the density perturbations 
grow under gravity, their distribution increasingly deviates from the initial Gaussian shape. If 
the density contrast $\delta$ had a perfect Gaussian distribution, with dispersion $\sigma$, there 
would always be a non-zero probability of having $\delta<-1$ in some region of space, which is 
clearly unphysical. Therefore, the real distribution of density perturbations will always be 
at least slightly non-Gaussian, with a cut-off at $\delta=-1$, i.e. positively skewed. Further, 
once the (initialy very rare) densest regions of the Universe start to turn-around and collapse, 
due to their own self-gravity, their density will increase much faster than that associated with 
most other regions at the same scale, which are still expanding with the Universe, i.e. evolving 
linearly. This leads to the development by the density distribution of a positive tail associated 
with high values for $\delta$, thus further increasing the skewness. These deviations do not matter 
as long as $\sigma\ll1$, for then the probability that $\delta<-1$ or the existence of regions in 
the process of turn-around is extremely small. Thus, in the early
stages of the gravitational evolution of non-correlated random density
perturbations it is a very reasonable assumption to take  their
distribution as perfectly Gaussian. However, as the value of $\sigma$
approaches 1, the  probability that $\delta<-1$ or of high values for
$\delta$ cannot be neglected any longer, and assuming the density
distribution to be Gaussian will induce significant errors in
calculations. 

In summary, gravitational evolution of density perturbations induces increasingly larger deviations 
from an initial Gaussian distribution, leading to a progressively more positively skewed 
distribution. Higher moments than the skewness, equally zero for a
Gaussian distribution, are also  generated in the process, though at
increasingly later times for the same amplitude. This means that among
these only the kurtosis (which compares the size of the side tails of
some distribution  against those of a Gaussian distribution) has developed
significantly by today on scales larger than  about 1 Mpc (on smaller
scales astrophysical processes irremediably mess up the calculations). 

In order to determine whether the initial probability distribution was Gaussian one then 
needs either to go to scales $R$ which are still evolving linearly, i.e. $\sigma(R)\ll1$, 
or one needs to quantify, using gravitational perturbation theory [see e.g. \cite{Bern}], 
the amount of skewness and kurtosis a Gaussian density field develops 
under gravitational instability for the values of $\sigma(R)$ observed.

If the values for the skewness and kurtosis were found to be well in excess of those 
expected, either the initial density distribution was non-Gaussian or 
structure did not form through gravitational instability. The problem is that we presently do not 
have direct access to the density field, though this will change in the near future by its 
reconstruction using data from large areas of the sky searched for gravitational lensed 
galaxies [see e.g. \cite{WH,Mel}]. Traditionally, the galaxy distribution 
has been used as a tracer of the density distribution, being assumed that the skewness and 
kurtosis of the density field should be the same as that of the galaxy field. However, it was soon realised 
that the galaxies could have a biased distribution with relation to the matter, i.e. 
$\delta_{gal}=b\delta$ where $b$ is the {\it bias parameter}. And a biased mass distribution with respect to 
galaxies closely resembles one which is unbiased, but which has a stronger degree of gravitational 
evolution. Dropping the assumption of linear bias, by considering the possibility that the galaxy 
distribution might depend on higher order terms of the density field, further complicates. Observationally 
the situation is also not very clear, with often incompatible values for both the skewness and kurtosis 
derived from the same galaxy surveys \cite{HG}. 

Another means of having access to the density field is by reconstructing it using the 
galaxy velocity field, under the assumption that galaxies move solely under the action of 
gravity. In fact, one can use directly the velocity field to constrain the initial distribution of 
the density perturbations, through its own skewness and kurtosis. The  
scaled skewness and kurtosis of the divergence of the velocity field have the advantage of 
not depending on a possible bias between the mass and galaxy distributions. But
they depend on the value of $\Omega_{0}$, in such a way that the velocity
field of a low-density Universe is similar to that of a high-density Universe
whose density field has evolved further under gravity. Nevertheless, by combining
measurements of both the scaled skewness and kurtosis it should in principle be possible to
determine if the initial density field was Gaussian \cite{Bern}.
However, only the line-of-sight velocity component is measurable,
hence reconstruction methods, like POTENT \cite{POT}, are used to recover the full 3D velocity 
field based solely on it. But such methods have their weaknesses, like
for example the need for very good distance indicators, which have up to today
hindered the use of the velocity field to determine whether the initial density
perturbations followed a Gaussian distribution.

The {\it topology} of the galaxy distribution can also be used to determine whether the initial density 
perturbations had a Gaussian distribution or not, if it is once more assumed that it reflects the 
underlying properties of the density field. The topological measure most 
widely used to distinguish between different underlying distributions is 
the {\it genus}, which essentially gives the number of holes minus the number of isolated 
regions, defined by a surface, plus one (e.g. it is zero 
for a sphere, one for a doughnut). Up to today all measures of the genus of the 
galaxy distribution seem compatible with it being Gaussian on the largest scales, 
above about $10\,h^{-1}$ Mpc  [see e.g. \cite{Can}], 
with  the prospects of even tighter constraints coming from the big galaxy surveys under way 
like the 2dF and the SDSS \cite{Colley}. 
 
\subsection{Scale-invariant perturbations ?}

The nearly scale-invariant nature of the initial perturbations,
as expected from inflation, can only be probed using large-scale structure data insofar
as one assumes a certain matter content in the Universe. 
This unfortunate situation results from the fact
that after horizon re-entry the subsequent evolution of the density
perturbations is greatly affected by both the type of matter present in the Universe
and the cosmic expansion  rate, which is in turn a function of the total
quantity of matter present in the Universe. 

The changes in the initial
power spectrum of density perturbations, which in our notation is given by 
$\delta_{H}^{2}(k)$, are encapsulated in the transfer function,
$T(k,t)$. Thus, unless one assumes a certain transfer function, we have 
no hope of recovering the shape of $\delta_{H}^{2}(k)$ from large-scale structure data. The
evidence for $\delta^{2}_{H}(k)\propto k^{n-1}$, as expected in the simplest inflationary models  
(in particular $n\leq1$), is therefore dependent on assumptions regarding the parameters that 
affect the transfer function. However, the choice of certain values for some of these 
parameters may come at a price, and be in conflict with observations not directly 
related with the power spectrum. For example, the observational data regarding light element
abundances implies that the present cosmic baryon density, $\Omega_{b}$, is about
$0.02\,h^{-2}$, with an error at 95 per cent confidence of less than about 20 per cent, if
homogeneous standard nucleosynthesis is assumed [see e.g. \cite{Tytler}]. 
Heavy tinkering with the value of $\Omega_{b}$ is tantamount to throwing away the standard 
nucleosynthesis calculation, thus requiring a viable alternative to be put in its place. It is 
also not possible to have any type of dark matter one might want to consider. If all the  dark 
matter particles had high intrinsic velocities, i.e. if they were hot, like neutrinos, then the
effect of free-streaming would completely erase all the perturbations on small scales,  e.g.
in the case of standard neutrinos on all scales smaller than about
$4\times10^{14}(\Omega_{0}/\Omega_{\nu})(\Omega_{\nu}h^{2})^{-2}\,{\rm M}_{\odot}$. Not
only it would then be impossible to reproduce the abundance of high-redshift objects, like
proto-galaxies, quasars, or damped Lyman-$\alpha$ systems, but the actual present-day 
abundance and distribution of galaxies would be radically different from that observed. It
seems very improbable that neutrinos presently contribute with more than about 30 per cent
of the total matter density in the Universe [see e.g. \cite{Viana,CHD,Novo}]. 

What is needed is to reformulate slightly the original question and ask instead
whether for the simplest, best observationally supported assumptions one can make, the
present-day power spectrum of density  perturbations is compatible with an initial
power-law shape, and in particular with one which is nearly scale-invariant. These 
assumptions are: the total energy density in the Universe is equal 
to or less than the critical density, i.e. $\Omega\leq1$, and results
from a matter component, plus a possible classical cosmological constant; 
the Hubble constant, in the form of $h$, has a value between 0.4 and 0.9; the baryon abundance, 
in the form of $\Omega_{b}h^{2}$, is within 0.015 to 0.025; the (non-baryonic) dark matter 
is essentially cold, with the possibility of any of the 3 known standard neutrinos having a 
cosmologically significant mass. Among all these assumptions, the one which could be more easily 
changed without conflicting with non-large scale structure data is the nature of the (non-baryonic) 
dark matter. There could be warm, decaying, or even self-interacting, dark matter, though usually 
the existence of some contribution by cold dark matter is found to be required to fit all the 
available large-scale structure data. 

In order to constrain the above free parameters in this {\it simplest} model, one needs at least
the same number of independent observational constraints. The most widely used are: 
the slope of the galaxy/cluster power spectrum; the present number density of rich galaxy clusters; 
the high-redshift abundance of proto-galaxies, quasars and damped Lyman-$\alpha$ systems; 
the amplitude of velocity bulk flows; the amplitude of CMBR temperature anisotropies, both on
large-angular scales, as measured by {\it COBE}, and on intermediate-angular scales, as presently 
measured by balloon experiments. While the first constraint directly limits the slope of the density power spectrum, the 
next four constraints do such only indirectly, by imposing possible intervals for the amplitude of 
the density power spectrum at specific scales. 

Other constraint that has started to be used recently, is the slope and normalisation of the density 
power spectrum on Mpc scales as inferred from the abundance and distribution of Lyman-$\alpha$ 
forest absorption features in the spectra of distant ($z\sim2.5$) quasars 
\cite{Croft}. However, these calculations depend on assuming 
a relatively simple physical picture for the formation of such features, being presently still 
unclear if such a picture provides a good approximation to reality. 

The comparison of the above defined {\it simplest} structure formation models with
CMBR anisotropy data is also presently not as clean as one would like. 
The two most important culprits are: the possibility  of a gravity wave contribution at the
{\it COBE} scale, thus allowing one to arbitrarily decrease the amplitude imposed by the
{\it COBE} result on the density power spectrum; the possibility of re-ionization,
which allows models with too much intermediate-scale power in the CMBR anisotropy angular
power spectrum to evade the observational limits on such scales. 

Again  taking refuge in the simplicity assumption, the {\it simplest} models 
would be those with a negligible contribution of gravity waves to the
large-angle CMBR anisotropy signal, together with no significant
re-ionization. In any case, it should be mentioned that relaxing these two further assumptions
does not open up a large region of parameter space \cite{Viana}. 

Unfortunately, the comparison of these {\it simplest} structure formation models with the 
observational constraints just described does not tell us much about the shape of the initial 
density power spectrum. The assumption of a power-law shape is perfectly compatible with the 
data, with the value of the spectral index being loosely constrained to be between 0.7 and
1.4 \cite{Viana,Novo}. However, by imposing further restrictions on the type of structure 
formation model considered, stronger constraints can be obtained. For example, 
if the Universe was Einstein-de Sitter then the value of the Hubble constant would have to be 
smaller than about 0.55, in order for the Universe to be more than 12 Gyr old. This does not matter 
much, because high values for $h$ increase the amount of small-scale power relative to large-scale 
power and at the same time suppress power at intermediate angular scales on the CMBR, and
for $h>0.55$  these two effects join together to exclude almost all viable Einstein-de
Sitter structure formation  models in the context of the {\it simplest} assumptions laid down
before \cite{Viana}. Restricting  ourselves to $0.5<h<0.55$, 
then yields a preferred value for the spectral index $n$ roughly between 0.9 and 1.1, with a total 
neutrino density of $\Omega_{\nu}\sim0.15$ [see e.g. \cite{Viana,GS,Novo}. Another example, is
the case of the $\Omega_{0}=0.3$ flat model, that preferred by the high-redshift type Ia 
supernova data of \cite{Perl}, for which values for the spectral index in excess of 1 tend
to be preferred \cite{Viana,Novo}. One should note however 
that both models are only marginally compatible with the observational data if $n=1.0$. 

Finally, it should be mentioned that, even within the simplified structure formation scenario we have
been assuming, there is enough room for initial density power
spectra with deviations from a power-law shape to be viable, as the survival of the
broken scale-invariance model testifies \cite{LPS}. This only goes to show the still scarceness of 
good quality large-scale structure data at present.

In the near future the Sloan Digital Sky Survey 
(SDSS) will allow a much better constraint to be imposed on the value of
$n$, by extending the measure of the slope of the galaxy power spectrum
to larger scales, thus probing a region of the power spectrum which has not  been in principle
too much affected by the dark matter properties, retaining therefore more 
information about the shape of the initial density power spectrum \cite{Love}. 

\section{The CMBR and inflation}

The measurement of the characteristics of the anisotropies in the temperature 
of the CMBR can teach us a lot not only about the processes that gave rise to the density 
perturbations responsible for the appearance of the large-scale structures we observe today, 
but also about cosmological parameters, like the Hubble constant and the matter and energy 
content in the Universe, and physical events that happened between decoupling and the present, 
like re-ionization and the formation of structure (e.g. cluster formation through the 
Sunyaev-Zel'dovich effect). However, there are so many possible ways in which the properties 
of the CMBR anisotropies can be altered, the most troublesome being by changing the nature of the 
initial perturbations and/or of the dark matter (energy), that in order to extract a tight 
interval of confidence for some parameter, e.g. the value of the Hubble constant, it is
necessary to make {\it a priori} some restrictions to the nature of the phenomena 
that might affect the characteristics of the CMBR temperature anisotropies.

Therefore, once again the strategy has to be to initialy work within a simplified framework, 
and only if the observational data demands it, expand the initial assumptions so as to widen the 
allowed parameter space. At present, the simplified framework within which most analysis of the 
present CMBR anisotropy data are made has 10 associated parameters, which may or may not be 
allowed to vary freely, and assumes primordial, adiabatic perturbations with a Gaussian 
distribution. The dark matter is taken to be of only two possible types, cold or massive 
standard neutrinos, and any dark energy is assumed to behave dynamically as a classical 
cosmological constant. We will follow the choice of parameters of 
\cite{TZ}, as slightly different, in practice equivalent, 
sets of parameters can be worked with. We then have 5 cosmological parameters: 
$\Omega_{k}$, the spatial curvature, equal to $-k/a^{2}H^{2}$;
$\Omega_{\Lambda}$, the energy density associated with a classical 
cosmological constant, $\Lambda$, which is equal to $\Lambda/3H^{2}$;
$w_{cdm}\equiv\Omega_{cdm}h^{2}$, the matter density in the form of 
cold dark matter; $w_{\nu}\equiv\Omega_{\nu}h^{2}$, the matter density in the form of 
standard neutrinos; $w_{b}\equiv\Omega_{b}h^{2}$, the matter density in the form of 
baryons. There are also 2 parameters that characterise the primordial power spectrum of the 
scalar adiabatic perturbations, which is assumed to have a power-law dependence with scale,
$\delta_{H}^{2}(k)\propto k^{n_{s}-1}$: $n_{s}$, the spectral index of the scalar perturbations; 
$A_{s}$, the contribution to the CMBR quadrupole anisotropy by scalar
perturbations, which is in practice equivalent to using $\delta_{H}$, 
the amplitude of the scalar perturbation power spectrum. And 2 parameters that characterise the 
primordial power spectrum of tensor perturbations, i.e. gravity  waves, which is also assumed to 
have a power-law dependence with scale, ${\it P}_{t}(k)\propto k^{n_{t}}$: 
$n_{t}$, the spectral index of the tensor perturbations; 
$A_{t}$, the contribution to the CMBR quadrupole anisotropy by tensor perturbations. Finally, 
there is 1 physical parameter: $\tau$, the optical depth to the surface of last scattering 
(equal to zero if there is no re-ionization after recombination occurred). From these quantities 
other well known parameters can be obtained, like the total matter density in the Universe, 
$\Omega_{m}=1-\Omega_{k}-\Omega_{\Lambda}$, and the value of the Hubble constant.

Again, it should be emphasised that any conclusions on the value of these parameters, obtained 
by comparing the theoretical expectation with the CMBR anisotropy data, is restricted to the 
tight framework imposed. The conclusions could be relaxed substantially if one was to consider 
introducing the added possibilities of part of the perturbations responsible for the CMBR 
anisotropies being active, incoherent, isocurvature, or with a non-Gaussian distribution, 
or the existence of more exotic types of dark matter (warm, or decaying, or 
self-interacting) or dark energy which does not behave dynamically as a cosmological
constant (e.g. an evolving scalar field). 

Through the study of the distribution of CMBR temperature anisotropies in the sky at 
different angular scales, it is possible to confirm or disprove the assumption of a Gaussian 
distribution for the perturbations. If these have a Gaussian distribution, them the anisotropies 
should also have one. At present only the {\it COBE} data has been searched for evidence of 
deviations from Gaussianity. Early results [see e.g. \cite{Kogut,GFE,Heavens}] 
were negative. However, more recently, there have been 
reports of strong evidence for deviations from a Gaussian distribution in the {\it COBE} data 
\cite{FMG,NFS,PVF,FGM,Mag}. However, some analysis continue to dispute the presence 
of a clear non-Gaussian signal \cite{BT,Barr,MHL}. 

Among the contradicting reports, it has become clear that the one-point distribution of the 
{\it COBE} anisotropies is indeed Gaussian, but the anisotropies seem not to be as randomly 
distributed across the sky as they should be under the Gaussian (random-phase) hypothesis, 
\cite{BT}. The later is supported by the fact 
that the statistical techniques that find evidence for non-Gaussianity seem to be those which 
are most sensitive to phase-correlations. The source for these could be instrument noise, the
method of extraction of the monopole, dipole or the Galactic contribution, unaccounted for 
systematic effects associated with the way {\it COBE} collected data \cite{BZG}, or of course real 
features in the microwave sky, which could be at the surface of last scattering or due to some 
unknown source of foreground contamination.  

Observational evidence for a primordial and adiabatic origin for the perturbations that produced 
the CMBR temperature anisotropies is scarce at present, even after the release of the Antarctica
long duration flight data from the {\it BOOMERanG} experiment \cite{deB,Lange} 
and data gathered by the first flight of the {\it MAXIMA} balloon \cite{Han}. 

The evidence for the primordial nature of the perturbations will come from the signature that the 
acoustic oscillations induced in the photon-baryon fluid leave in the angular power spectrum of the 
CMBR anisotropies: a sequence of peaks in $\ell$ space. Their adiabatic nature is revealed by
the spacing between the peaks: $\ell_{1}:\ell_{2}:\ell_{3}...$ corresponds to ratios of 
$1:2:3:...$. Isocurvature primordial perturbations also generate a sequence of peaks in the 
CMBR temperature anisotropy angular power spectrum, but at different locations. The first 
peak due to the acoustic oscillations, generated by the potential wells that result from the 
isocurvature primordial perturbations after they enter the horizon, is moved to an higher $\ell$ 
value, and the ratios between the first three (real) peak locations are now $3:5:7:...$. The ratios 
between the peaks in both cases only depend weakly on the assumed cosmological parameters, and 
therefore provide an excellent test of the adiabatic or isocurvature nature of the fluctuations 
\cite{HWa,HWb}. Unfortunately, at present, though the {\it BOOMERanG LDF} and the {\it MAXIMA-1} 
temperature anisotropy measurements go to sufficiently high $\ell$ to probe the expected location of a 
second peak in the adiabatic case, the evidence is inconclusive. Nevertheless, purely isocurvature 
perturbations seem to be highly disfavoured just by the shape of the first peak \cite{KV}. So, 
presently one cannot use 
solely CMBR anisotropy data to distinguish between an adiabatic or an isocurvature origin for the
density perturbations. In any case, the present known location of the first acoustic peak
($\ell\sim200$) is sufficient to say that if the Universe is spatially flat then most probably 
the initial perturbations were adiabatic, as one then expects $\ell_{1}\simeq220$. 
Isocurvature perturbations necessarily need the Universe to be closed, which moves the
horizon scale at decoupling to larger angular scales, given that if the Universe was flat
then the first acoustic peak for primordial isocurvature perturbations would be expected at
$\ell_{1}\simeq340$. If one admits the possibility that both adiabatic and isocurvature modes are 
present, then even with {\it MAP} it will be difficult to estimate the relative amplitude of both. 
One will have to wait for {\it Planck} to measure such a quantity with less than 10 per cent error 
\cite{BMT}.

As discussed before, the appearance of acoustic 
peaks in the angular power spectrum of the CMBR temperature anisotropies is not 
proof of an inflationary (or primordial) origin for the initial perturbations, though their 
absence could only be (realistically) explained by early and substantial 
(i.e. {\large $\tau$} close to  unity) re-ionization of the intergalactic
medium. However, the fact that at least one acoustic peak has been detected argues
against the later scenario, and its presence is evidence for some degree of coherence
of the initial perturbations (which does not mean their are primordial, as previously
mentioned). 

In summary, the primordial, adiabatic and Gaussian nature of the initial perturbations
seems to be still a viable first assumption, providing the simplest coherent framework within 
which it is possible to explain the present CMBR temperature anisotropy data. Unfortunately, 
even under this framework the CMBR data presently does not tell us much about the parameters 
$n_{s}$, $A_{s}$,$n_{t}$, and $A_{t}$, which characterise the power spectrum of the scalar 
and tensor perturbations in the {\it simplest} inflationary models. 

Gravitational waves can only contribute significantly to the CMBR temperature
anisotropies on scales above $1^{\circ}$, or $\ell<100$, given that at the surface of last 
scattering any gravitational  waves inside the horizon had their amplitude severely 
diminished by redshifting. This means that only the {\it COBE} and {\it Tenerife} data 
\cite{TZ} might in principle be sensitive to the tensor spectral index $n_{t}$. However, the
effect of varying $n_{t}$ can be easily masked by simultaneously varying the scalar
spectral index $n_{s}$, or by reducing the tensor contribution to the large-angle
CMBR anisotropy to levels small enough that (almost) any variation of $n_{t}$
could be accommodated within the measured errors. Therefore, at present there are
no useful observational limits on the value of $n_{t}$. 

The ratio between $A_{t}$ and $A_{s}$, also sometimes called $T/S$,
should in principle be a more readily measurable quantity, as it affects the height of
the first acoustic peak relatively to the large-angle Sachs-Wolfe plateau, and also 
the CMBR anisotropy normalisation of the density perturbation power spectrum, 
thus affecting the formation of large-scale structure. By analysing these two effects, 
\cite{ZSW} concluded that 
values of $T/S$ as large 4 are still viable, as long as $n_{s}$ is simultaneously
allowed to go up to 1.2. Adding the initial data coming from the {\it BOOMERanG} and 
{\it MAXIMA} experiments seems to decrease the maximum possible value for $T/S$ to 2 \cite{KMR}. 
Restricting $n_{s}$ to be lower or equal to unity, given that 
this is what is predicted in the simplest inflationary models that produce a 
cosmologically significant background of gravitational waves, then the ratio $T/S$ would need 
to be lower than about 1. In particular, in the case of power-law inflation models the 
observational data seems to require $T/S<0.5$.

Finally, $n_{s}$ is also not very well constrained
at present by CMBR anisotropy data alone or in conjunction with structure formation data. 
The CMBR data by itself badly constrains $n_{s}$ if all the 10 parameters
associated with the simplest inflationary motivated structure formation models are allowed 
to vary \cite{TZ}. By fixing the matter density in the form of baryons to be 
$\Omega_{b}h^{2}=0.02$ (as implied by observations if homogeneous standard nucleosynthesis is 
assumed), and a reasonable range for the Hubble constant, $0.55<h<0.75$, a interval for 
$n_{s}$ can be obtained, going from 0.8 to about 1.5 \cite{TZ}. However, 
values for $n_{s}$ higher than 1.3 can only be obtained at the cost of having a very large 
value for $T/S$ \cite{TZ,LeDour}. And in any case the the initial data from the {\it BOOMERanG} and 
{\it MAXIMA} experiments seem to disfavour values for $n_{s}$ above 1.2 \cite{Huetal,Jaffe,KMR}. 
We have seen that structure formation observations do not constrain very well $n_{s}$ either, with values 
between 0.7 and 1.4 being possible. Joining all the data together would thus indicate that a value for $n_{s}$ 
close to 1 is preferable, with the possibility of going as low as 0.8 or as high as 1.3. The upper 
limit is also confirmed by bounds imposed on the production of primordial black holes 
\cite{GL}. 

Looking into the future, the prospects of measuring the inflationary parameters $n_{s}$, $A_{s}$, 
$n_{t}$, and $A_{t}$ from CMBR anisotropies are mixed. Data from balloon experiments like 
{\it BOOMERanG} and {\it MAXIMA} will not help much in improving 
these constraints, except for a better handle on the value of $n_{s}$. As mentioned, initial data from the 
two experiments has already further restricted the value of $n_{s}$ to being within about 10 per cent of 
unity \cite{Huetal,Jaffe}. The results from the ${\it MAP}$ 
and ${\it Planck\,Surveyor}$ satellites will however change the situation. All studies that have 
been done to date of the precision with which both ${\it MAP}$ and ${\it Planck}$ will be able to 
estimate different cosmological parameters use the so-called {\it Fisher information matrix}. This 
gives the manner in which experimental data depends  upon a set of underlying theoretical parameters 
that one wishes to measure. Within this set of parameters, the Fisher matrix yields a lower limit 
to error bars and hence an upper limit on the information that can be extracted from such a data 
set, i.e. the Fisher matrix reveals the best possible statistical error bars achievable from 
an experiment. With the further assumption of Gaussian-distributed signal and noise, Fisher 
matrices can be constructed from the specifications of CMBR experiments. However, given that 
the Fisher matrix formalism in practice asks how well an experiment can distinguish the true 
model of the Universe from other possible models, the results for the error bars on the parameters 
of these models will depend on which {\it true} model is chosen as input. Several groups have 
tried to estimate how well ${\it MAP}$ and ${\it Planck}$ will be able to extract information 
on the cosmological parameters from CMBR anisotropies using the Fisher matrix formalism 
[e.g. \cite{BET,ZSS,WSS}]. I will quote the results from 
the work by \cite{EHT}, who were  
probably those who more thoroughly explored, within the {\it simplest} model context described 
before, the parameter degeneracies that may appear when the information is extracted from the 
CMBR data. They even included the possibility of a dependence with scale of the spectral index of 
density perturbations [see also \cite{CGL}]. Regarding  $n_{s}$, \cite{EHT} conclude that among 
the four inflationary quantities it is the most readily measurable: just by using the CMBR temperature 
anisotropy angular power spectrum it will be possible to estimate it with an error of about 0.1 
($1\sigma$) through {\it MAP} and around 0.05 for {\it Planck}. Also using the CMBR polarisation 
information the error tends to be halved in the case of {\it MAP} and reduced by a factor of 5 
for {\it Planck}. The other quantities, $A_{s}$, $n_{t}$, and $A_{t}$, will not be much better known 
after {\it MAP} and {\it Planck}, if just the CMBR temperature anisotropy spectrum is considered. Only
by including polarisation information it will be possible to vastly improve the
knowledge we have about the values taken by $T/S$ and $n_{t}$. 

The CMBR is expected to be partially linearly polarised [see \cite{HW97} for a thorough review]. 
The polarisation signal can be decomposed into two separate and orthogonal components, the so-called 
{\it E} and {\it B} modes, with each mode having their own associated power spectra, which will 
depend in a different way on cosmological parameters. A cross-correlation signal 
between the temperature anisotropy and {\it E}-mode polarisation maps will also be present 
[see e.g. \cite{Kam,Kin}]. The knowledge of the polarisation signal will clearly contribute to 
improving the CMBR constraints on cosmological and perturbation parameters, but not equally 
on all of them. Those parameters that most gain from the use of the CMBR polarisation 
information are the ratio $T/S$, $n_{t}$ and the optical depth {\large $\tau$}. The reason is that only 
vector or tensor perturbations can give rise to {\it B}-mode polarisation, which can also arise from 
re-ionization of the intergalactic medium. However, vector perturbations 
rapidly decay in an expanding Universe, if all perturbations are primordial (e.g of inflationary 
origin). Consequently, while the total polarisation signal can help in the better estimation of all
parameters, the power spectrum information associated with the {\it B}-mode
polarisation will help particularly in bringing down the {\it relative} 
errors in the estimation of $T/S$, $n_{t}$ and {\large $\tau$}, most noticeably if the first and 
last are small \cite{Kin,EHT}. However, from a practical point 
of view, it will not be easy to disentangle the CMBR polarisation signal from foreground 
polarisation sources, given the  present lack of knowledge about their nature \cite{PSB}.  

Interestingly, the polarisation signal can also provide one of the strongest tests known 
of the inflationary paradigm. Because polarisation is generated at the last scattering surface, 
models in which perturbations are causally produced, necessarily on sub-horizon scales, 
cannot generate a polarisation signal on angular scales larger than about $2^{o}$, approximately 
the horizon size at photon-baryon decoupling \cite{SZ}. Hence, 
polarisation correlations on such scales would indicate either a
seemingly acausal mechanism for the generation of the perturbations or
re-ionization at work. However, the two mechanisms can in principle be
disentangled due to their different $\ell$ dependence on large-angular scales 
\cite{Zal}. If what seems like acausality at work is 
proven, then it has been argued \cite{Liddle} that inflation 
provides the only mechanism through which it can be generated, by expanding initialy sub-horizon 
quantum fluctuations to super-horizon sizes. Unless it is postulated the existence of super-horizon 
perturbations since the begining of the Universe, which would give rise to a new initial 
conditions problem. 

The measurement of the polarisation signal, by improving the estimation of $T/S$ and $n_{t}$, 
will also provide the possibility of checking whether the so-called inflationary consistency 
relation holds, $T/S\simeq-7n_{t}$ 
[e.g. \cite{LLPR}]. If this is shown than 
single-field slow-roll models will receive a tremendous boost. Note, however, that there are 
inflationary models which do not predict such a relation. 

Finally, we have until know only discussed what future constraints can be imposed on the 
cosmological parameters and the nature of the perturbations solely through the CMBR temperature 
anisotropy and polarisation signals. Clearly it would help if some of the degeneracies 
inherent to CMBR analysis could be broken by using large-scale structure data and 
direct measurements of the geometry and expansion rate of the Universe. The most 
promising large-scale structure data to be expected in the near future is the SDSS data, 
which together with the already well known local rich cluster abundance, provides 
a means of constraining the slope and normalisation of the present-day density power spectrum. 
Supernova type Ia and direct measurements of the Hubble constant will enable independent 
estimates of respectively the geometry and expansion rate of the Universe. Further, the 
evolution with redshift of the number density of rich galaxy clusters is a powerful method 
for determining the present total matter content, $\Omega_{0}$, while gravitational lensing 
is able to impose interesting constraints on the value of a possible cosmological constant. 

\section{Direct detection of gravitational waves}

Gravitons are the propagating modes associated with transverse, traceless 
tensor metric perturbations, and they behave as a superposition of two 
minimally coupled scalar fields, each corresponding to a polarisation state. 
As a result, the graviton field, which is massless, has a spectrum of quantum 
mechanical fluctuations similar to the one obtained for the scalar field 
$\phi$. For each polarisation state, the {\it rms} amplitude of the tensor metric 
perturbations associated with a given Fourier mode at horizon crossing 
is then \cite{KT}

\begin{eqnarray}
\left(\Delta h\right)^{2}_{k}&\equiv&\frac{V}{2\pi^{2}}\,k^{3}\,|h_{k}|^{2}
=\frac{4}{\pi}\left(\frac{H}{m_{{\rm Pl}}}\right)^{2}\nonumber\,,
\end{eqnarray}

where $V$ is the volume associated with the Fourier expansion, and the value of $H$ is to be 
calculated when the comoving scale $k$ crosses outside the Hubble radius during inflation. 
The phases of the Fourier modes $h_{k}$ are again independent and randomly distributed. 
Consequently, the power spectrum at horizon re-entry, ${\it P}_{t}(k)$, 
contains all the information necessary to describe the stochastic background 
of gravitational waves generated during inflation. Again, in general, 
${\it P}_{t}(k)$ can be approximated as a single power-law, 

$$
{\it P}_{t}(k)\propto k^{n_{t}}\nonumber\,.
$$

A scale-invariant power spectrum for the tensor perturbations then corresponds to $n_{t}=0$. 
For the simplest single-field inflationary models, as long as the gravitational waves are
produced during the slow-roll phase, necessarily $n_{t}\leq0$. Further, under such
conditions, and in the simplest models, the ratio between ${\it P}_{g}(k)$ and
$\delta^{2}_{H}(k)$ is approximately equal to
$-n_{t}/2$, which is another way of stating the so-called {\it inflationary consistency relation} 
\cite{LLPR}. Note that in these models, 
if the expansion of the Universe during inflation is perfectly exponential, one gets 
$n_{t}=0$, but then the amplitude of the tensor perturbation also effectively tends to zero compared
with the amplitude of the (equally scale-invariant) scalar perturbation. However, in
some slow-roll models with more complicated potentials, like intermediate inflation, it is 
possible to have an important tensor contribution to the perturbation spectrum at horizon re-entry 
though the power spectrum of density perturbations is scale-invariant. 

If a gravitational wave background produced during inflation was detected using the CMBR, 
and it was possible to estimate both the amplitude and spectral index of the tensor perturbation 
contribution to the CMBR temperature anisotropy and polarisation signals, the next step would be to 
try to detect locally those gravitational waves. If successful, such a detection would 
do much the same for the credibility of inflation as was achieved for the Hot Big Bang theory 
itself by the detection of the CMBR.  

The simplest quantity to compare with the experimental sensitivity of gravitational wave detectors
is the present-day contribution per logarithmic interval of the gravitational waves to the total
energy density, 

$$
\Omega_{gw}(k)\equiv\frac{1}{\rho_{c}}\frac{d\,\rho_{gw}(k)}{d\,\log k}\,,
$$

where $\rho_{c}=3H_{0}^{2}/8\pi G$ and $\rho_{gw}$ is the energy density of the stochastic background
of gravitational waves with comoving wavenumber $k$. Given that 

$$
\Omega_{gw}(k)=\frac{1}{6}\left(\frac{k}{H_{0}}\right)^{2}\left(\Delta h\right)^{2}_{k}\,,
$$

we then get in the case of a initial scale-invariant power spectrum of gravitational waves 

$$
\Omega_{gw}(k)=\frac{2}{3\pi}\left(\frac{k}{H_{0}}\right)^{2}\left(\frac{H}{m_{{\rm Pl}}}\right)^{2}\,.
$$

The shape of the initial power spectrum is broken at the scale of matter-radiation equality, 
$k_{eq}=6.22\times10^{-2}\,\Omega_{0}h^{2}\,\sqrt{3.36/g_{*}}\,{\rm Mpc}^{-1}$, where $g_{*}$ 
is the effective number of relativistic degrees of freedom (equals 3.36 for the standard cosmology 
with 3 massless neutrino species), as gravitational waves that enter the horizon prior to 
matter-radiation equality redshift more slowly with time. Given that we are only interested in
gravitational waves that can be detected locally, we will concentrate on those which entered 
the horizon during radiation domination. These correspond to $k\ll2\times10^{-24}{\rm m}$, 
or $f\ll10^{-16}\,{\rm Hz}$, where $f=c\,k/2\pi$ is the frequency.

Working within the simplest inflationary models, those which obey the consistency relation 
$T/S\simeq-7{\large n_{t}}$, and requiring the CMBR temperature anisotropies detected on 
large-angular scales by {\it COBE} to be reproduced, one obtains \cite{Turner} 

\begin{eqnarray}
\Omega_{gw}(k)&=&5.1\times10^{-15}\,h^{-2}\,\left(\frac{g_{*}}{3.36}\right)
\left(\frac{\large n_{t}}{\large n_{t}-1/7}\right)
\times \exp\left({\large n_{t}}\ln\frac{k}{cH_{0}}\right)\nonumber\,.
\end{eqnarray}

If the power spectrum of tensor perturbations is not a perfect power-law there will be 
small corrections to this expression, which will be in principle more important for the 
smallest scales, i.e. high $k$. A sensitivity of $\Omega_{gw}(k)\,h^{2}\sim10^{-15}$ is therefore needed 
for a serious search for local gravitational waves produced during inflation. With its initial strain 
detectors, the Earth-based {\it LIGO} (Laser Interferometer Gravitational Wave Observatory) should be 
able to identify a  stochastic background of gravitational waves provided $\Omega_{gw}h^{2}$ is at
least a few times $10^{-3}$, at its most sensitive operating frequency of roughly 100 Hz, with the
limit dramatically improving by possibly 6 orders of magnitude with more advanced strain detectors
installed in a later phase \cite{Magg}. Unfortunately, this misses the mark 
by six orders of magnitude. 

Because the energy density in gravitational waves is proportional to the rms strain $\Delta h$ squared
times frequency squared, a detector operating at lower frequency has better 
energy-density sensitivity for fixed strain sensitivity. Earth-based detectors cannot operate at 
frequencies below about 10 Hz because of seismic noise, but space-based operators can. The 
{\it LISA} (Laser Interferometer Space Antenna) mission has already been approved by ESA, with an  
initialy predicted launch date for only around 2020, but which may be brought down to later this 
decade if NASA gets interested in a joint effort. It will have a peak sensitivity in terms of
$\Omega_{gw}h^{2}$ of about $10^{-12}$ at a frequency close to $10^{-3}$ Hz \cite{Magg}, 
which is more promising, but still misses by at least three 
orders of magnitude the required sensitivity level for the detection of a local stochastic background 
of inflationary produced gravitational waves. Also, at frequencies above $10^{-4}$ Hz it is expected 
that the stochastic background of gravitational waves produced by compact white-dwarf binaries will 
swamp the inflationary signal. However, {\it LISA} may be able to disentangle the two backgrounds, given 
that it rotates in orbit and so it will be sensitive at different times to regions in the Galactic 
plane, where the binaries are, and outside \cite{Magg}. 

\section{Is the Universe flat ?}

One of the problems of standard cosmology is the near flatness of the Universe. For the present total
energy density in the Universe to be within one order of magnitude of the critical density, i.e.
$\Omega_{total}\sim1$ today, at the Planck time ($10^{-43}$ s) the value of $\Omega_{total}$ could not
deviate from unity by more than $10^{-60}$. This is one of the problems that can be solved by assuming
the existence of an inflationary period in the very early Universe. The simplest models of inflation
predict that $\Omega_{total}$ should in practice be equal to unity today, i.e. the Universe to be 
spatially flat. Therefore, these models would receive strong support if it was shown that presently 
$\Omega_{total}=1$. Note that because there are inflationary models which predict the Universe not to be 
presently flat, then proof that $\Omega_{total}$ is different from unity today would not disprove the 
inflationary paradigm, but simply be evidence that if inflation indeed occurred than it did so in a more 
complicated fashion than it is generally assumed. 

Recently, tantalising evidence has appeared that seem to indicate that the Universe is flat.  
There are several methods that directly or indirectly probe the geometry of the Universe. Those which
presently provide the cleanest constraints on the geometry are the position of the first acoustic peak 
on the CMBR temperature anisotropy angular power spectrum and the magnitude-distance relation for
Supernovae type Ia. Two other methods provide limits essentially on the total amount of non-relativistic
matter in the Universe, $\Omega_{m}$, the evolution with redshift of the abundance of rich galaxy
clusters and deviations from Gaussianity measured either through the galaxy or the cluster velocity
fields. Finally, the number of observed gravitational lensed high-redshift objects puts limits mainly 
on the possible contribution to the total matter density by a classical cosmological constant,
$\Omega_{\Lambda}$.  Given that most analysis assume only these two possible contributions to the total
energy density in  the Universe, non-relativistic matter, $p\simeq0$, and  a cosmological
constant, $p=-\rho$, I will not consider other eventual contributions with different
equations of state, for example arising from an evolving scalar field. 

Let me then summarise what we presently know about $\Omega_{total}$, $\Omega_{m}$ and 
$\Omega_{\Lambda}$. This list does not pretend to be exhaustive, as some of the results 
cannot be expressed through a simple function of $\Omega_{m}$ and $\Omega_{\Lambda}$. Some of the
limits were determined by \cite{RHa} from results in the references given. 

From the angular scale of the first acoustic peak in the CMBR anisotropy spectrum, 
under the assumption of Gaussian adiabatic initial perturbations: $\Omega_{total}>0.85$ \cite{Huetal}; 
$\Omega_{total}=1.15\pm0.20$ \cite{Lange1}; $\Omega_{total}=0.90\pm0.15$ \cite{Balbi}; 
$\Omega_{total}=1.11\pm0.07$ \cite{Jaffe}. From the magnitude-distance relation for Supernovae type Ia: 
$0.8\Omega_{m}-0.6\Omega_{\Lambda}=-0.2\pm0.1$ \cite{Perl}. From the cluster abundance evolution with redshift, 
under the assumption of Gaussian initial perturbations: $\Omega_{m}=0.2^{+0.3}_{-0.1}$ \cite{BF};  
$\Omega_{m}=0.45\pm0.20$ \cite{Eke}; $\Omega_{m}>0.3$ \cite{VL}; 
$\Omega_{m}=0.45\pm0.10$ \cite{Henry}; $\Omega_{m}=0.75\pm0.20$ \cite{Blan}. From the cosmic velocity field, 
$\Omega_{m}>0.3$ \cite{Dekel} at more than 95 per cent confidence from 
the amplitude of diverging flows of galaxies from voids \cite{DR} and from the skewness 
of the velocity field assuming the initial density distribution to be Gaussian \cite{ND}. From the 
gravitational lensing of objects at high-redshift:  
$\Omega_{\Lambda}=0.70\pm0.16$ \cite{CY}; $\Omega_{m}>0.26$ \cite{FKM}; 
$\Omega_{m}<0.62$ \cite{Cooray}; $-1.78<\Omega_{\Lambda}-\Omega_{m}<0.27$ \cite{Helbig}.

The quoted results indicate that the situation is still 
too confusing for one to be able to say with any degree of certainty which is the value 
of either $\Omega_{m}$ or $\Omega_{\Lambda}$. However, it seems clear that the 
best explanation for the combined data is an Universe which is spatially flat \cite{RHa,RHb}.

\end{document}